\documentstyle[12pt]{article}

\newcommand{\EQ}{\begin{equation}}
\newcommand{\EN}{\end{equation}}
\newcommand{\bea}{\begin{eqnarray}}
\newcommand{\ena}{\end{eqnarray}}

\renewcommand{\a}{\alpha}

\renewcommand{\d}{\delta}

\newcommand{\la}{\lambda}

\newcommand{\shalf}{\frac{1}{2}}
\newcommand{\pa}{\partial}
\begin{document}

\topmargin 0pt
\oddsidemargin 5mm

\renewcommand{\Im}{{\rm Im}\,}
\newcommand{\NP}[1]{Nucl.\ Phys.\ {\bf #1}}
\newcommand{\PL}[1]{Phys.\ Lett.\ {\bf #1}}
\newcommand{\NC}[1]{Nuovo Cimento {\bf #1}}
\newcommand{\CMP}[1]{Comm.\ Math.\ Phys.\ {\bf #1}}
\newcommand{\PR}[1]{Phys.\ Rev.\ {\bf #1}}
\newcommand{\PRL}[1]{Phys.\ Rev.\ Lett.\ {\bf #1}}
\newcommand{\PTPS}[1]{Prog.\ Theor.\ Phys.\ Suppl.\ {\bf #1}}
\newcommand{\MPL}[1]{Mod.\ Phys.\ Lett.\ {\bf #1}}
\newcommand{\IJMP}[1]{Int.\ Jour.\ Mod.\ Phys.\ {\bf #1}}
\newcommand{\JP}[1]{Jour.\ Phys.\ {\bf #1}}
\renewcommand{\thefootnote}{\fnsymbol{footnote}}
\begin{titlepage}
\setcounter{page}{0}
\vspace{0.5cm}
\rightline{OS-GE 25-92}

\vspace{2cm}
\begin{center}
{\Large Interactions of Discrete States with Nonzero Ghost Number \\
in $c=1$ $2D$ Gravity}
\vspace{2cm}

{\large Nobuyoshi Ohta and Hisao Suzuki} \\
\vspace{1cm}
{\em Institute of Physics, College of General Education, Osaka
University, \\ Toyonaka, Osaka 560, Japan} \\
\end{center}
\vspace{2cm}
\centerline{{\bf{Abstract}}}

We study the interactions of the discrete states with nonzero ghost
number in $c=1$ two-dimensional ($2D$) quantum gravity. By using
the vertex operator representations, it is shown that their interactions
are given by the structure constants of the group of the area preserving
diffeomorphism similar to those of vanishing ghost number. The
effective action for these states is also worked out. The result suggests
the whole system has a BRST-like symmetry.

\vspace{4cm}
\end{titlepage}
\newpage
\setcounter{footnote}{0}

Much attention has recently been paid to nonperturbative treatment of
two-dimensional ($2D$) quantum gravity in terms of the matrix model
\cite{NON,PGM}. Most of the remarkable results there are now understood
in the continuum approach using the techniques in the conformal field
theory~\cite{DDK,SPA}. There still remain many problems to be clarified,
however, in order to understand the full theory. The problem with which
we are mainly concerned in this paper is the interaction of the so-called
``discrete states" in the $c=1$ conformal field theory coupled to $2D$
gravity~[5-9].

In the conformal gauge, the $c=1$ quantum gravity may be regarded
effectively  as a string theory in two dimensions with suitable
background charge. Thus it is expected that there is only the degree
of freedom corresponding to the ``center of mass" or ``tachyon", since
there are no transverse directions. However, it has been found that
there exist other discrete degrees of freedom in the $c\leq 1$ theory
coupled to the $2D$ quantum gravity, both in the matrix models and
in the Liouville approach [5-9].

The nature of these ``discrete states" is most effectively studied
in the Liouville theory. By using the BRST formulation of the $c=1$
quantum gravity, all the physical states characterized by the BRST
cohomology have been enumerated and it has been found that there
are indeed an infinite number of physical states with ghost number
$N_{FP}=0,\pm 1$ at the discrete values of momenta~[6-11].
These states may be interpreted as higher string states~\cite{POL}, but
they exist only for fixed values of momenta, allowing for no
usual particle interpretation.

Recently it has been pointed out that the dynamics of these discrete
states are governed by the symmetry group of the area preserving
diffeomorphism~[10-15].
For $c=1$, these states with $N_{FP}=0$ are known to form
representations of the $SU(2)$ Kac-Moody algebra~[10-13,15,16].
By using the vertex operator representation
of these states by means of the $SU(2)$ algebra, Klebanov and Polyakov
have computed the three point interactions and have proposed an
effective action for these discrete states~\cite{KPO}.
This action is further made complete by including the scalar degrees
of freedom~\cite{MST}. However, the interactions involving
those extra states with $N_{FP}\neq 0$ have not been studied and
their role in the theory remains elusive. It is thus interesting to
examine their interactions and try to clarify their role in order
to understand the whole structure of the theory.

The purpose of this paper is to fill the above gap and, in particular,
determine the three point interactions of the extra states with
$N_{FP}\neq 0$. We will do this using the  vertex operator
representation of these states given for the $c=1$ theory~\cite{IO2}.

Denoting the $c=1$ matter and Liouville fields by $X(z)$ and
$\phi(z)$, respectively, one has the energy-momentum tensor
\EQ
T(z) = -\shalf(\pa X)^2 -\shalf(\pa\phi)^2+\sqrt{2}\pa^2\phi.
\EN
The discrete physical
state are characterized by the $SU(2)$ current algebra defined by
\EQ
J^{\pm}(z)=e^{\pm i\sqrt{2}X(z)}, \hspace{1cm}
J^0(z)= \frac{1}{\sqrt{2}} i\pa X(z).
\EN
The states with $N_{FP}=0$ are obtained by acting the following
operators on the physical vacuum $|0> \equiv |0>_{X,\phi}\otimes
c_1|0>_{bc}$:
\EQ
\Psi^{(\pm)}_{Jm}(z)= \sqrt{\frac{(J+m)!}{(2J)!(J-m)!}}(J^-_0)^{J-m}
 e^{i\sqrt{2}JX(z) +\sqrt{2}(1\pm J)\phi(z)},
\EN
where $J_0^{-}$ is the contour integral over the current $J^{-}$ around
$z$ and $J=1,2, \cdots ; m=-J,-J+1, \cdots, J-1, J$.\footnote{For
simplicity, we consider integer spin $J$ in this paper.} These generate
spin $J$ multiplet with  $N_{FP}=0$. Their interactions are now known
to be described by the symmetry
of the area preserving diffeomorphism~\cite{KPO,MST}.

The extra states with ${\rm N_{FP}}\not=0$ also fall into representations
of the $SU(2)$ current algebra. It has been shown that these states are
generated by the following operators~\cite{IO2}
\bea
{\tilde \Psi}^{(-)}_{J-1,m}(z) &\sim& (J^-_0)^{J-m-1} \oint_z \frac{dw}
{2\pi i} \frac{{\hat c}(w)J^-(w)}{w-z} e^{i\sqrt{2}JX(z)
 +\sqrt{2}(1+ J)\phi(z)}, \\
{\tilde \Psi}^{(+)}_{J-1,m}(z) &\sim& (J^-_0)^{J-m-1} \oint_z \frac{dw}
{2\pi i} b(w)e^{-iX(w)/\sqrt{2}-\phi(w)/\sqrt{2}} e^{i\sqrt{2}(J-1/2)
 X(z) +\sqrt{2}(3/2- J)\phi(z)},
\ena
where the caret on the ghost field $c(w)$ in eq. (4) means that
the zero mode ${c_0}$ is removed. As has been noted in ref.~\cite{IO2},
this is necessary to make the created states in the ``relative
cohomology"\footnote{In other words, we subtract the states proportional
to $\pa c(z)$ in the ``absolute cohomology" which are in spin $J$
representation.} and these states form spin $(J-1)$ representation.

Applying the operator (4) on the physical vacuum, we easily see that
these create states with $N_{FP}=1$
\bea
&&{\tilde \Psi}^{(-)}_{J-1, m}(0)|0>\otimes c_1|0>_{bc} \nonumber\\
&\sim& (J^-_0)^{J-m-1} \oint_0 \frac{dw}{2\pi i} \sum_{n\neq 0} c_n w^{-n}
 e^{-i\sqrt{2}X(w)} |p_X=\sqrt{2}(J-1), p_\phi=-\sqrt{2}iJ>, \nonumber\\
&=& (J^-_0)^{J-m-1} \sum_{n\geq 1}c_{-n}S_{2J-1-n}\left(-\frac{\sqrt{2}
 \a^X_{-n}}{n}\right)|p_X=\sqrt{2}(J-1), p_\phi=-\sqrt{2}iJ>,
\ena
in agreement with ref.~\cite{BMP1}. Here we have used the Schur
polynomial defined by
\EQ
\exp \left( \sum_{k\geq 0}x_kz^k \right)=\sum_{k\geq 0}S_k(x)z^k.
\EN
We can show that other states created by (3)-(5) also agree with
those given in ref.~\cite{BMP1} and that these states are BRST
invariant.

In order to examine the interactions of these states, we first
follow ref.~\cite{KPO} and compute the operator product expansions
(OPEs). From the ghost number conservation and the dependence
on the zero modes of $X$ and $\phi$, we should have
\bea
{\tilde \Psi}^{(+)}_{J_1-1,m_1}(z){\tilde \Psi}^{(-)}_{J_1+J_2-2,-m_1-m_2}(0)
&=& \cdots +\frac{1}{z}\sum F_{J_1-1,m_1,J_1+J_2-2,-m_1-m_2}^{J_2,-m_2}
 \Psi^{(-)}_{J_2,-m_2}(0)+\cdots , \nonumber\\
F_{J_1-1,m_1,J_1+J_2-2,-m_1-m_2}^{J_2,-m_2} &=&
C_{J_1-1,m_1,J_1+J_2-2,-m_1-m_2}^{J_2,-m_2}g(J_1,J_2),
\ena
where $C$ are the Clebsch-Gordan coefficients and $g(J_1,J_2)$ is an
unknown function to be determined. For $J_3=J_1+J_2-1,m_3=-m_1-m_2$,
we have
$$
C_{J_1-1,m_1,J_3-1,m_3}^{J_2,-m_2} =
\frac{(-1)^{J_1-1-m_1}N(J_3-1,m_3)}{N(J_1-1,m_1)N(J_2,m_2)}
 [m_1J_2-m_2(J_1-1)], \eqno(9a)
$$
$$
N(J,m) = \sqrt{\frac{(J-m)!(J+m)!}{(2J-1)!}}. \eqno(9b)
$$
\setcounter{equation}{9}
Notice that ${\tilde \Psi}^{(\pm)}_{J-1,m}$ have spin $(J-1)$. We
thus see that the last factor in eq. (9a) is the structure constant of
the area preserving diffeomorphism~\cite{WIN}.

Our next task is to determine $g(J_1,J_2)$. For this purpose,
we consider the special case $m_1=-J_1+2$ and $m_2=-J_2$. One finds

\bea
&&\sqrt{2(J_1-1)}{\tilde \Psi}^{(+)}_{J_1-1,-J_1+2}(z)
{\tilde \Psi}^{(-)}_{J_1+J_2-2,J_1+J_2-2}(0) \nonumber\\
&=& \oint_z\frac{d\xi}{2\pi i}\oint_z\frac{d\zeta}{2\pi i}
   \oint_0\frac{d\eta}{2\pi i}(\xi-\eta)(\xi-z)^{-2J_1+1}(\xi-\eta)^{-2}
  \xi^{2J_1+2J_2-2}(\zeta-\eta)^{-2}, \nonumber\\
&&\times (\zeta-z)^{2-2J_1}\zeta^{2J_1+2J_2-3}(z-\eta)^{2J_1-1}
  \eta^{-2J_1-2J_2+3}z^{-2J_2-1}\Psi^{(-)}_{J_2,J_2}(0), \nonumber\\
&=& \frac{1}{z} \oint_1\frac{dw}{2\pi i}\oint_1\frac{dv}{2\pi i}
   \oint_0\frac{du}{2\pi i}(w-v)(w-1)^{-2J_1+1}(w-u)^{-2}
  w^{2J_1+2J_2-2}(v-u)^{-2} \nonumber\\
&&\times (v-1)^{2-2J_1}v^{2J_1+2J_2-3}(1-u)^{2J_1-1}
  u^{-2J_1-2J_2+3}\Psi^{(-)}_{J_2,J_2}(0),
\ena
where we have used the fact that the contraction $<b(z)\hat c(w)>\sim
w^2/z^2(z-w)$ because the zero mode $c_0$ is removed from $c(w)$.
The $u$-integration is then deformed to be winding around $u=v$ and $w$.
We thus get
\bea
&& -\frac{1}{z} \oint_1\frac{dw}{2\pi i}\oint_1\frac{dv}{2\pi i}
  \frac{1}{(w-v)}[(1-w)^{2J_1-1}w^{-2J_1-2J_2+3}
  -(1-v)^{2J_1-1}v^{-2J_1-2J_2+3}] \nonumber\\
&&\times \frac{\pa^2}{\pa w\pa v}[(w-v)(w-1)^{-2J_1+1}w^{2J_1+2J_2-2}
 (v-1)^{-2J_1+2}v^{2J_1+2J_2-3}\Psi^{(-)}_{J_2,J_2}(0) \nonumber\\
&& = -\frac{(2J_1+2J_2-3)!}{(2J_1-3)!(2J_2-1)!}
  \Psi^{(-)}_{J_2,J_2}(0).
\ena
Combining eq. (11) with (8) and (9), we find
\EQ
F=\frac{(-1)^{J_1-1-m_1}N(J_3-1,m_3)}{N(J_1-1,m_1)N(J_2,m_2)}
 \frac{(2J_3-1)!}{(2J_1-3)!(2J_2-1)!\sqrt{2J_2(J_1-1)(J_3-1)}}
[m_1J_2-m_2(J_1-1)].
\EN

In principle, it should be possible to compute other OPEs, but
we have not been able to complete it. Here, instead of pursuing this
line, let us directly compute the three point functions involving
these states. This is equivalent to computing all possible OPEs at once.
This also serves as a check of our above result.

It is easy to see that the only nonvanishing function is
\EQ
<0|\Psi^{(+)}_{J_2,m_2}(z_1)c(z_1){\tilde \Psi}^{(+)}_{J_1-1,m_1}(z_2)
c(z_2){\tilde \Psi}^{(-)}_{J_1+J_2-2,-m_1-m_2}(0)c(0)|0>,
\EN
which will be a constant that has the same structure as the coefficient
$F$ in eq. (8). There is no other coupling of those states with
``tachyonic" states.

To compute this, we again specialize to the case $m_1=-J_1+1$
and $m_2=-J_2+1$. First we compute (13) with zero mode $c_0$ included
and then subtract the contribution from $c_0$. The first contribution
(times $\sqrt{2J_2}$) is given by
\bea
&&-\oint_{z_1}\frac{dw}{2\pi i}\oint_{z_2}\frac{dv}{2\pi i}
   \oint_0\frac{du}{2\pi i}(z_1-z_2)^{2J_1+J_2-2}z_1^{-2J_1+1}z_2^{-2J_2}
 (w-z_1)^{-2J_2}(z_1-v)^{-2J_2} \nonumber\\
&& \times (z_1-u)^{2J_2+1}(w-v)(w-z_2)^{-2J_1+1}(w-u)^{-2}
 w^{2J_1+2J_2-2}(v-z_2)^{-2J_1+1} \nonumber\\
&& \times (v-u)^{-2}v^{2J_1+2J_2-2}(z_2-u)^{2J_1} u^{-2J_1-2J_2+2}.
\ena
At first sight, this appears to be $SL_2$ non-invariant and hence
depends on $z_1$ and $z_2$. We will see that actually this gives a
constant.\footnote{This can be understood from the fact that (14) is
made $SL_2$ invariant by multiplying with
$\lim_{Z\to \infty}\frac{Z-u}{Z}=1$.}

The contour integration over $u$ is again deformed and one finds only
the contribution form $u=\infty$. The result turns out to be
\bea
&&-\oint_{z_1}\frac{dw}{2\pi i}\oint_{z_2}\frac{dv}{2\pi i}
   (z_1-z_2)^{2J_1+J_2-2}z_1^{-2J_1+1}z_2^{-2J_2}(w-v)(w-z_1)^{-2J_2}
 (w-z_2)^{-2J_1+1} \nonumber\\
&&\times w^{2J_1+2J_2-2}(z_1-v)^{-2J_2}(v-z_2)^{-2J_1+1}v^{2J_1+2J_2-2}.
\ena
We then perform the $w$-integration, which is deformed to be around
$w=z_2$ and $\infty$. The contribution from $w=z_2$ drops out
because of symmetry and we are left with
\EQ
- \oint_{z_2}\frac{dv}{2\pi i} [2J_2z+(2J_1-1)z_2-v]
 (z_1-v)^{-2J_2}(v-z_2)^{-2J_1+1}v^{2J_1+2J_2-2}.
\EN
Performing the $v$ integration, we finally get
\EQ
-\frac{(2J_1+2J_2-2)!}{(2J_1-2)!(2J_2-1)!},
\EN
which is independent of $z_1$ and $z_2$.

The contribution from the zero-mode term, on the other hand, can be
computed similarly. Using the correlator of ghosts
\EQ
<0|c(z_1)b(v)c(z_2)c_0uc(0)|0>
=\frac{uz_1^2z_2^2(z_1-z_2)}{v^2(z_1-v)(v-z_2)},
\EN
we find it is given by
\bea
&&\oint_{z_2}\frac{dw}{2\pi i}\oint_{z_2}\frac{dv}{2\pi i}
   \oint_0\frac{du}{2\pi i}(z_1-z_2)^{2J_1+J_2-2}z_1^{-2J_1+2}z_2^{-2J_2+1}
 (w-z_1)^{-2J_2}(z_1-v)^{-2J_2} \nonumber\\
&& \times (z_1-u)^{2J_2}(w-v)(w-z_2)^{-2J_1+1}(w-u)^{-2}
 w^{2J_1+2J_2-2}(v-z_2)^{-2J_1+1}(v-u)^{-1} \nonumber\\
&& \times v^{2J_1+2J_2-3}(z_2-u)^{2J_1-1} u^{-2J_1-2J_2+2},
\ena
which is $SL_2$ invariant. By a similar procedure, one finds
\EQ
- 2J_2 \frac{(2J_1+2J_2-3)!}{(2J_1-2)!(2J_2-1)!},
\EN
Hence the three point correlation (12) for $m_1=-J_1+1$ and
$m_2=-J_2+1$, which is given by the difference of (17) and (20), is
\EQ
-\frac{(2J_1+2J_2-3)!}{(2J_1-3)!(2J_2-1)!},
\EN
Comparing this with the general structure of the correlation,
we find it is given precisely by (12).

If we redefine the fields by
\bea
\Psi^{(\pm)'}_{J,m}(z) &=& -\left[ N(J,m)(2J-1)!\sqrt{\frac{J}{2}}
  \right]^{\pm 1} \Psi^{(\pm)}_{J,m}(z), \nonumber\\
{\tilde \Psi}^{(+)'}_{J-1,m}(z) &=& (-1)^{J-m}N(J-1,m)(2J-3)!
 \sqrt{\frac{J-1}{2}} {\tilde \Psi}^{(+)}_{J-1,m}(z), \nonumber\\
{\tilde\Psi}^{(-)'}_{J-1,m}(z) &=& \left[ N(J-1,m)(2J-3)!(2J-1)
  \sqrt{\frac{J-1}{2}} \right]^{-1}{\tilde \Psi}^{(-)}_{J-1,m}(z),
\ena
we get
\EQ
<0|\Psi^{(+)'}_{J_2,m_2}(z_1)c(z_1){\tilde \Psi}^{(+)'}_{J_1-1,m_1}(z_2)
c(z_2){\tilde \Psi}^{(-)'}_{J_1+J_2-2,-m_1-m_2}(0)c(0)|0>
=m_1J_2-m_2(J_2-1),
\EN
Therefore, introducing variables $g^{(s),A}_{J,m}$ and ${\tilde g}
^{(s),A}_{J-1,m}$ $(s=\pm)$ for these states, the effective action for
the three-point interactions is given by
\EQ
S_{3,gh}=-g_0\sum_{J_1,m_1,J_2,m_2,A,B,C} [J_2m_1-(J_1-1)m_2]f^{ABC}
{\tilde g}^{(-),A}_{J_1+J_2-2,-m_1-m_2}{\tilde g}^{(+),B}_{J_1-1,m_1}
 g^{(+),C}_{J_2,m_2}\int d\phi,
\EN
where we have introduced the open string coupling constant $g_0$ and
the Chan-Paton index $A$ in the adjoint representation of a Lie group.
This can be rewritten in terms of a field defined as
\bea
\Phi(\phi,\theta,\varphi) &=& \sum_{s,A,J,m}T^Ag^{(s)A}_{J,m}M^s(J,m)
 D^J_{m,0}(\varphi,\theta,0)e^{(sJ-1)\phi}, \nonumber\\
 {\tilde \Phi}^{(\pm)}(\phi,\theta,\varphi) &=& \sum_{A,J,m}T^A
 {\tilde g}^{(\pm)A}_{J,m}M^{\pm}(J,m) D^{J}_{m,0}
 (\varphi,\theta,0) e^{(\pm J-1)\phi},
\ena
where $M^s$ are normalization constants defined by
\EQ
M^+(J,m)=\frac{(J-1)!}{\sqrt{(2J-1)!}}N(J,m), \hspace{1cm}
M^-(J,m)=\frac{(-1)^m}{4\pi}\frac{(2J+1)\sqrt{(2J-1)!}}{(J-1)!N(J,m)}.
\EN
Note that the fields with $N_{NF}\neq 0$ have opposite statistics to
those with $N_{FP}=0$.
Using the Poisson brackets for the rotation matrix~\cite{KPO}
\bea
&&\{D^{J_1}_{m_1 0},D^{J_2}_{m_2 0}\} \nonumber\\
&=& i\frac{N(J_3,m_3)}{N(J_1,m_1) N(J_2,m_2)}
 \sqrt{\frac{(2J_1-1)!(2J_2-1)!}{(2J_3-1)!}}\frac{(J_3-1)!}
 {(J_1-1)!(J_2-1)!}(J_2m_1-J_1m_2)D^{J_3}_{m_3,0}, \nonumber\\
\ena
we finally obtain
\EQ
S^{(1)}_3=-2ig_0 \int d\phi e^{2\phi}\int_{S^2}d^2x
 \varepsilon^{ij} \mbox{Tr}\left({\tilde \Phi}^{(-)}
\frac{\pa{\tilde \Phi}^{(+)}}{\pa x^i} \frac{\pa \Phi}{\pa x^j}\right),
\EN
where $x^i=(\theta, \varphi)$. This action is essentially identical
to that for the states with $N_{FP}=0$.

What is the physical meaning of this result? The form of the effective
action reminds us the similar structure in the BRST formulation of the
nonabelian gauge theory~\cite{KUO}. Here similarly we suspect that the
whole theory may have a BRST-like invariance in the target space,
just as in the string field theory.
In fact, we can see the symmetry in the present cubic action. For this
purpose, it is convenient to write the action for the states without
ghost number in terms of the fields
\EQ
\Phi^{(\pm)}(\phi,\theta,\varphi) = \sum_{A,J,m}T^Ag^{(\pm)A}_{J,m}
M^\pm(J,m) D^J_{m,0}(\varphi,\theta,0)e^{(\pm J-1)\phi}.
\EN
The total action for the cubic terms is then
\EQ
S_3=ig_0\int d\phi e^{2\phi}\int_{S^2}d^2x
 \varepsilon^{ij} \mbox{Tr}\left( \Phi^{(-)}
\frac{\pa \Phi^{(+)}} {\pa x^i} \frac{\pa \Phi^{(+)}}{\pa x^j}
 - 2{\tilde \Phi}^{(-)} \frac{\pa{\tilde \Phi}^{(+)}} {\pa x^i}
 \frac{\pa \Phi^{(+)}}{\pa x^j} \right),
\EN
which is invariant under
\bea
\d\Phi^{(+)} &=& \la {\tilde \Phi}^{(+)}, ~~
\d{\tilde\Phi}^{(-)} = \la \Phi^{(-)},\nonumber\\
\d{\tilde\Phi}^{(+)} &=& 0, ~~ \d\Phi^{(-)} = 0.
\ena
Note that this is a nilpotent transformation. This symmetry is similar
to the transformation generated by the charge
\EQ
Q \sim \oint\frac{dz}{2\pi i}b(z)e^{-(iX(z)+\phi(z))/\sqrt{2}},
\EN
which is the operator to create the states with ghost number from those
without it (see eq.~(5)), except that this changes the spins of the
states (for example, $Q \Psi^{(+)}_{J,m}={\tilde \Psi}^{(+)}_{J-1/2,
m-1/2}$). The action can be written as
\EQ
S_3=ig_0\int d\phi e^{2\phi}\int_{S^2}d^2x
 \varepsilon^{ij} \d\left[\mbox{Tr}\left( {\tilde \Phi}^{(-)}
 \frac{\pa \Phi^{(+)}}{\pa x^i} \frac{\pa \Phi^{(+)}}{\pa x^j}
 \right)\right],
\EN

It is then natural to conjecture that these ``ghost degrees of freedom"
play the role of canceling part of the contribution from the $N_{FP}=0$
states, just as the Faddeev-Popov ghosts. Indeed, Bershadsky and
Klebanov~\cite{BEK} have computed the
one-loop partition function in $c=1$ gravity and found that it contains
the contribution only from the primary fields of the form
$e^{i(pX+{\bar p}{\bar X})}$ (those for $|m|= J$ and ``tachyons")
because the contribution of each special primary field with $|m|\leq
J-1$ cancels with that of the descendant of the previous one. To really
check this possibility in our approach, we have to examine the one-loop
contribution by using the effective action.

We hope that our finding that the ``ghost states" have nonvanishing
correlations and that they have the same structure as that of the
$N_{FP}=0$ states will help to get further insight into the theory.

\vspace{1cm}
\noindent{\em Acknowledgement}

We would like to thank N. Sakai and Y. Tanii for useful discussions.


\newpage
\begin{thebibliography}{99}
\bibitem{NON} E. Br\'{e}zin and V. Kazakov, \PL{B236} (1990) 144;
              M. R. Douglas and S. Shenker, \NP{B335} (1990) 635;
              D. J. Gross and A. A. Migdal, \PRL{64} (1990) 127;
	      \NP{B340} (1990) 333.
\bibitem{PGM} G. Parisi, \PL{B238} (1990) 209;
              D. J. Gross and N. Miljkovi\'{c}, \PL{B238} (1990) 217;
              E. Br\'{e}zin, V. Kazakov and Al. B. Zamolodchikov,
              \NP{B338} (1990) 637;
              P. Ginsparg and J. Zinn-Justin, \PL{B240} (1990) 333;
	      J. Ambj{\o}rn, J. Jurkiewicz and A. Krzywicki, \PL{B243}
	      (1990) 209, 213.
\bibitem{DDK} J. Distler and H. Kawai, \NP{B321} (1989) 509;
              F. David, \MPL{A3} (1989) 1651.
\bibitem{SPA} N. Seiberg, \PTPS{102} (1991) 319;
	      J. Polchinski, in {\em Strings '90}, ed. R. Arnowitt et al.
	      (World Scientific, Singapore, 1991) p.62; \NP{B357} (1991) 241.
\bibitem{GKN} D. J. Gross and I. Klebanov, \NP{B344} (1990) 475;\\
              D. J. Gross, I. Klebanov and M. Newmann, \NP{B350} (1990)333.
\bibitem{POL} A. M. Polyakov, \MPL{A6} (1991) 635.
\bibitem{LZ} B. H. Lian and G. J. Zuckerman, \PL{B254} (1991) 417;
              \PL{B266} (1991) 21.
\bibitem{BMP1} P. Bouwknegt, J. M. McCarthy and K. Pilch, CERN preprints,
              CERN-TH.6162/91 (1991); TH.6279/91 (1991).
\bibitem{IO1} K. Itoh and N. Ohta, Fermilab preprint,
              FERMILAB-PUB-91/228-T (1991), to appear in \NP{B}.
\bibitem{WIT} E. Witten, \NP{B373} (1992) 187.
\bibitem{KPO} I. Klebanov and A. M. Polyakov, \MPL{A6} (1991) 3273.
\bibitem{JRO} D. Kutasov, E. Martinec and N. Seiberg, Rutgers preprint,
	      RU-91-49 (1991).
\bibitem{IO2} K. Itoh and N. Ohta, Osaka preprint, OS-GE 22-91 (1991),
	      to appear in \PTPS (1992).
\bibitem{BMP2} P. Bouwknegt, J. M. McCarthy and K. Pilch, CERN preprint,
              CERN-TH.6346/91 (1991).
\bibitem{MST} Y. Matsumura, N. Sakai and Y. Tanii, Tokyo Inst. of Tech.
	      preprint, TIT/HEP-186 (1992).
\bibitem{ITO} K. Itoh, Texas A \& M preprint, CTP-TAMU-42/91 (1991).
\bibitem{WIN} I. Bakas, \PL{B228} (1989) 57;\\
	      C. Pope, L. Romans and X. Sen, \NP{B339} (1990) 191;\\
	      E. Bergshoeff, M. P. Blencowe and K. S. Stelle, \CMP{128}
	      (1990) 213.
\bibitem{KUO} T. Kugo and I. Ojima, \PTPS{66} (1979) 1.
\bibitem{BEK} M. Bershadsky and I. R. Klebanov, \NP{B360} (1991) 559.
\end{thebibliography}
\end{document}